# A Lightweight Signature-Based IDS for IoT Environment

**Nazim Uddin Sheikh[1,2]**, **Hasina Rahman[1]**, **Shashwat Vikram[2] and Hamed AlQahtani[1]**
Macquarie University[1], Sydney, Australia
Institute of Engineering & Management[2], Kolkata, India

**Abstract**
With the advent of large-scale heterogeneous networks comes the problem of unified network control resulting in security lapses that could have otherwise avoided. A mechanism is needed to detect and deflect intruders to safeguard resource constraint edge devices and networks as well. In this paper we demonstrate the use of an optimized pattern recognition algorithm to detect such attacks. Furthermore, we propose an Intrusion Detection System (IDS) methodology and design architecture for Internet of Things that makes the use of this search algorithm to thwart various security breaches. Numerical results are presented from tests conducted with the aid of NSL KDD cup dataset showing the efficacy the IDS.

*Key words: false positive; NSL KDD dataset; internet of things, IDS.*

## 1. Introduction

Intrusion detection is a process of identifying vulnerability in the network. One of the integral parts of cyber security system is intrusion detection method that has drawn a significant attention since a couple of decades. IDS can successfully implement and manage observed network security controls [1]. IDS implements intrusion detection functionality. Intrusion Detection & Prevention System (IDPS) is a powerful tool which is being used to detect, deflect or in some manner counteract attempts at unauthorized access. These systems execute an increasingly imperative role in the field of network security to prevent interlopers. Many IDPS tools have been designed, among them are Snort [2], Eye Bros [3], and Hawkeye [4]. However, the false alarm issue remains a challenging problem to solve. An intrusion detection system's efficacy can severely be influenced by noise. Malicious packets generated from software bugs, corrupt DNS data, and local packets that escaped might produce a substantial number of false alarms. The actual quantity of intrusions is often far lower than the number of false alarms that the real intrusions are often failed to detect [5].

Designing an absolute IDS (i.e. free from false alarm) is far away from reality. There are three common detection methodologies which are as follows: Signature-Based detection, Statistical Anomaly-Based detection and Stateful Protocol Analysis [6]. IDS can be considered as an effective and efficient security technology. Its functionalities include detection, prevention and probably react to the illegitimate access of resources or someone trying to penetrate into the system by breaching the security policies [7]. IDS deploys many techniques in order to safeguard critical systems [8]. The signature-based detection approach is also known as misuse detection. The misuse detection approach analyses network activities for known attacks usually through string-searching algorithms. Pattern matching algorithms have been used in the designing of IDS. On the contrary, statistical anomaly detection technique makes its decision based on a record of normal network or system behavior, generally created with the help of machine learning techniques. There are many anomaly-based detection processes that exist along with their commercial applications [9, 10]. Both of these above discussed approaches have certain pros and cons. The signature-based approach has general tradition of very low false positive alarm rates compared to others.

## 2. Related Work

In this section, we will describe a popular string matching algorithms used in the context of intrusion detection system.

We have elucidated few string matching algorithms used in context with intrusion detection system. With the advent of the Internet era and its easy accessibility has brought a concern of unwanted and illicit penetration of security-wall of different organizations and government agencies. To soothe this concern, many open-source and commercial intrusion detection and prevention systems are designed. Some of the existing IDPSs which are described as follows.

### 1.1 Snort (NIDS)

Snort is a lightweight, cross-platform, network sniffing tool and developed as a full-featured network intrusion detection system (NIDS) [13] developed by Martin Roesch in 1990 in order to detect attacks targeting his home network. Snort is efficient, misuse based and open-source IDS. It generates alarms based on predefined misuse rule-set. It makes use of tcpdump-formatted files to capture packets coming from the networks [2]. The Snort search engine is driven by Aho-Corasic multi-pattern matching algorithm. However, a typical single-threaded base engine architecture of Snort IDS confines its performance.

Snort is comprised of following components:
- Packet Decoder

The packet-capturing engine of Snort utilizes libpcap library. Captured packets are processed by packet decoder engine and decoded packets become acquiescent with the network-layer protocols.
- Preprocessors/ Input Plug-ins
- Detection Engine
- Logging/Alerting System
- Log Files

There are many advantages of Snort. Some of its advantages are as follows.
(i) Snort is an Open Source IDS, thus free to use.
(ii) It is flexible to customize it in different environments and it is also compatible with operating systems like Windows and Linux.
(iii) It supports auto-updating rules and could be utilized in decentralized mode.

## 1.2 Suricata

The idea behind the design and development of Suricata IDS [14] was to replace Snort. It was engineered by Open Information Security Foundation (OISF), financed by the US Department of Homeland Security. Unlike Snort, Suricata implements multi-threaded engine to improve its efficacy compared to Snort. However, the approach of multi-threaded architecture may lead to a slower detection rate in actual practice. Suricata has some additional unique features like the capability of detecting common network protocols even though they are not operating over standard ports assigned to them.

## 1.3 Bro IDS

Bro is Open Source network intrusion detection (NIDS) framework and comprises of multi-level modular architecture underlying network layer in the ISO-OSI seven-layer model [3]. It is a platform independent framework. Bro passively monitors inbound packets and sniffs for malicious activities. The activities like multi-layer analysis, policy imposition, behavioral controlling, and policy-oriented detection are conducted by Bro and it observes attacks after elicitation of network traffic into proper semantic format in order to execute the comparison engine. There are many significant advantages of using Bro IDS:
- It effectively captures data from Gbps networks and can perform with great efficacy in high-speed environment.
- More sophisticated and complex signatures are constructed by using this framework.
- It has flexibility to customize features.

However, Bro has some limitations like
- It is difficult and more time consuming to deploy.
- It does not have any Graphical User-Interface (GUI) and is compatible only with Linux OS.

## 3. Features of the Dataset

The KDD Cup 99 [16] dataset has been assembled and processed by the Cyber Systems and Technology Group [17] of MIT Lincoln Laboratory, under Defense Advanced Research Projects Agency and Air Force Research Laboratory (AFRL), United States in the year 1999 for the evaluation of computer network intrusion detection systems. The complete dataset contains about five million session records and each record refers to a TCP/IP connection that is composed if 41 features. These 41 features are both qualitative and quantitative in nature [18]. These features are grouped into following four classes [19].

### 3.1.1 Basic Features:
The first six consecutive features in the dataset fall under this section that are directly accumulated from packet headers.

### 3.1.2 Content Features:
To access data section in the TCP packets, domain knowledge is applied. For instance, number of failed login attempts is an example of content features.

### 3.1.3 Time- Based Traffic Features:
Features that are fashioned in order to capture properties of network traffic that last for more than 2 second temporal window and is divided into two subcategories which are "same host" features and "same service" features.

### 3.1.4 Host-Based Traffic Features:
The ports or hosts are being scanned by some probing attacks taking place over a much larger time interval. For example, features like destination host were designed with the help of a window of 100 connections or records to the same host despite a time window [20].

The dataset contains four major types of attacks [21] which are as follows: Denial of Services (DoS), Remote to Local (R2L), User to Root (U2R), and Probe with 22 types of sub-categories of attacks.

## 4. Proposed Methodology

In this section we have discussed our proposed IDS architecture and the functionalities of its each component. The proposed intrusion detection system consists of four components. The four-layered architecture of the intrusion detection system with its schematic diagram is depicted in figure 1.

### 4.1 Signature Generator (Layer 1):

The main functionality of the engine is to generate signatures of different attack types from the training/ labelled dataset. In other words, the labelled dataset is translated into corresponding nucleotide sequences and are stored in the signature database attack-wise. The signature database must be updated time to time in order to meet the real-time requirements.

### 4.2 Pattern Generator (Layer 2):

The engine in the layer 2 that processes the sessions of the testing or unlabeled dataset (accumulated from the packet headers and payloads from the network traffics) and then convert each of those features into corresponding DNA sequence and finally storing them in a temporary database.

### 4.3 Intrusion Detection Engine (Layer 3):

Intrusion Detection Engine (IDE) is the most intrinsic part of the system. The engine is comprised of a novel pattern matching algorithm [24] using some fixed real values (Relative Frequency of English Letters) in conjunction with asset of arithmetic operations that analyze and compare the session DNA pattern with the signatures

### 4.4 Output Engine (Layer 4):

The post-analysis step is to generate various log files through which the administrator will be apprised of the status of each session specifying/alerting whether it is normal-type or attack-type.

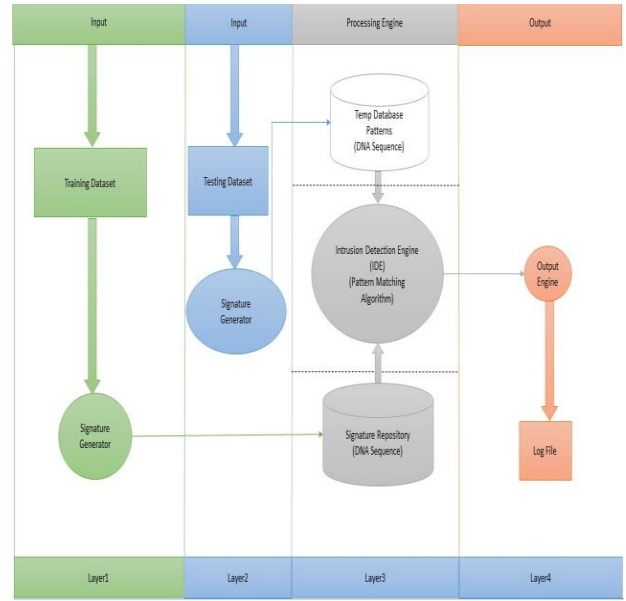

**Figure1. A Four-Layered Schematic Diagram of Intrusion Detection System**

## 5. Experimental Evaluation

In this section we examine the performance of the envisaged intrusion detection methodology. In particular, we evaluate the false positive and false negative occurrences in the experiments. Extensive experiments were done on NSL KDD dataset. Table 1 shows the occurrences of false alarms: false positives and false negatives both in every thousand records in the testing dataset.

TABLE-I

| Number of Samples | False Positives |
|---:|---:|
| 10,000 | 9 |
| 20,000 | 26 |
| 30,000 | 54 |
| 40,000 | 81 |
| 50,000 | 86 |
| 60,000 | 102 |
| 70,000 | 111 |
| 80,000 | 123 |
| 90,000 | 148 |
| 100,000 | 156 |

The occurrence of false positives in different ranges of traffics derived from the experiment is shown in figure 2.

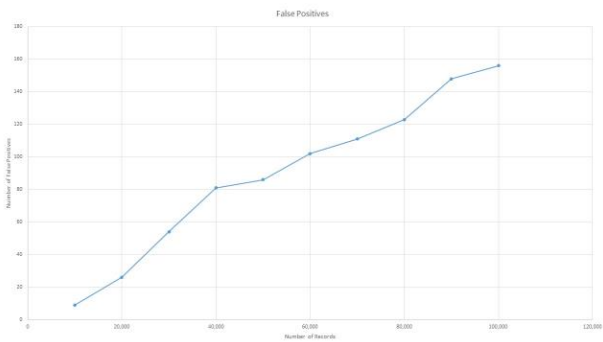

Figure 2. False positives

The graph represents the number false positives that are encountered by the IDS. Further it shows that the number of mismatches increase with the rise of the number of strings. The above graph demonstrates that the number of false positives increase with increase of number of samples in the training dataset.

## 6. Conclusion and Future Directions

In this paper, we studied state of the art of intrusion detection systems. Furthermore, we envisaged a signature-based intrusion detection system using a fast pattern matching algorithm which outperforms in detecting known attacks. The simulation result of our IDS portrays an optimistic scope for future research. The occurrence of false alarms are shown in the graphs. Our continuous endeavor hopefully will lead us to transit this off-line IDS into a real-time efficient intrusion detection and prevention system (IDPS) that will not only detect network traffics with aberrant nature but also safeguard the networks from the suspicious activities that penetrate into the networks.

### Acknowledgments

The main segment of this research work was done at Institute of Engineering & Management, Kolkata, India.

## References

[1]. Bejtlich, Richard, The Tao of Network Security Monitoring: Beyond Intrusion Detection, Addison-Wesley, 2004
[2]. M. Roesch ''Snort–Lightweight Intrusion Detection for Networks,'' 13th Systems Administration Conference, USENIX, 1999.
[3]. Bro IDS. Project WWW Page. http://http://www.bro-ids.org/, 2011.
[4]. Pantos "Packet Reading with libpcap", April 2010 http://www.systhread.net/texts/200805lpcap1.php
[5]. Anderson, Ross (2001). *Security Engineering: A Guide to Building Dependable Distributed Systems*. New York: John Wiley & Sons. pp. 387–388. ISBN 978-0-471-38922-4.
[6]. Scarfone, K. and Mell, P. (2007) Guide to Intrusion and Prevention System (IDPS). Department of Commerce, National Institute of Standard and Technology, Technology Administration.
[7]. Gowadia, V., Farkas, C., and Valtorta, M., Paid: A probabilistic agent-based intrusion detection system. *Journal of Computers and Security,* 2005.
[8]. Rebecca Base and Peter Mell, *NIST Special Publication on Intrusion Detection Systems*. Infidel, Inc., Scotts Valley, CA and National Institute of Standards and Technology, 2001.
[9]. S. Staniford, J. Hoagland, and J. McAlerney. Practical automated detection of stealthy portscans. Journal of Computer Security, 10(1-2):105-126, 2002.
[10]. M. Tavallaee, E. Bagheri, W. Lu and A. Ghorbani, "A Detailed Analysis of the KDD'99 CUP Data Set", The 2nd IEEE Symposium on Computational Intelligence Conference for Security and Defense Applications (CISDA), **(2009)**.
[11]. R. M. Karp and M. Robin, "efficient Randomized Pattern Matching Algorithms." Center for Research in Computing Technology, Harvard University, Report TR- 31-81, 1981.
[12]. Alfred V. Aho and Margaret J. Corasik. Efficient String Matching: An Aid to Bibliographic. Search: Bell Labs, Communications of the ACM Jun 1975 Volume 18 Number 6.
[13]. J. Koziol, "Intrusion Detection with Snort", Sams Publishing, May 2003.
[14]. E. Albin, "A Comparative Analysis of Snort And Suricata Intrusion Detection Systems", Naval Postgraduate School, Dudley Know Library, September 2011.
[15]. Lewand, R. Cryptological Mathematics. Washington, DC: Mathematical Association of America, 2000.
[16]. R.P. Lippmann, D.J. Fried, I. Graf, J.W. Haines, K.R. Kendall, D. McClung, D. Weber, S.E. Webster, D. Wyschogrod, R.K. Cunningham et al., *Evaluating Intrusion Detection Systems: the 1998 DARPA Off-line Intrusion Detection Evaluation.* Proc. DARPA Information Survivability Conference and Exposition, 2000, vol. 2.
[17]. KDD Cup 99 Data., http://kdd.ics.uci.edu/databases/kddcup99/kddcup99.html [Last access: 12th November 2008].
[18]. S. Stolfo, W. Fan, W. Lee, A. Prodomidis and R.K. Chan, *Cost-based Modeling for Fraud and Intrusion Detection: Results from the JAM Project* in DARPA Information Survivability Conf., 2000
[19]. Khaled Labib**.** 2004. Computer Security and Intrusion Detection**.** ACM Crossroads. Volume 11(1): 2.
[20]. Khaled Labib**.** 2004. Computer Security and Intrusion Detection**.** ACM Crossroads. Volume 11(1): 2.
[21]. J. H. Güneş Kayacýk, A. Nur Zincir-Heywood, Malcolm I. Heywood. Selecting Features for Intrusion Detection: A Feature Relevance Analysis on KDD 99 Intrusion Detection Datasets**.**
[22]. M. Tavallaee, E. Bagheri, W. Lu, and A. Ghorbani, "A Detailed Analysis of the KDD CUP 99 Data Set," *Submitted to Second IEEE Symposium on Computational Intelligence for Security and Defense Applications (CISDA)*, 2009.
[23]. J. McHugh, "Testing intrusion detection systems: a critique of the 1998 and 1999 darpa intrusion detection system evaluations as performed by lincoln laboratory," *ACM Transactions on Information and System Security*, vol. 3, no. 4, pp. 262–294, 2000.
[24]. N. Sheikh, K. Mustafi, I. Mukhopadhyay, "A Unique Approach to Design an Intrusion Detection System using an Innovative String Matching Algorithm & DNA Sequence" 2016 IEEE 7[th] Annual Ubiquitous Computing, Electronics & Mobile Communication Conference.